\documentclass[preprint,12pt]{elsarticle}

\usepackage{amsmath,amssymb,amsfonts}
 \usepackage{hyperref}
\usepackage{booktabs}
\usepackage{mathrsfs}
\usepackage{footnote}
\newcommand{\tabincell}[2]{\begin{tabular}{@{}#1@{}}#2\end{tabular}}
\begin{document}

\begin{frontmatter}

\title{When Healthcare Meets Off-the-Shelf WiFi: A Non-Wearable and Low-Costs Approach for In-Home Monitoring\tnoteref{mytitlenote}}
\tnotetext[mytitlenote]{This work is supported by National Key R\&D Program of China under
Grant 2018YFC0810204 and National Natural Science Foundation of China under Grant 61671073.}
%\author{Lingchao Guo\fnref{myfootnote}}
\author[mymainaddress,mysecondaryaddress,mythirdaddress]{ Lingchao Guo }
\ead{rita\_guo@bupt.edu.cn}
\author[mymainaddress,mysecondaryaddress,mythirdaddress]{ Zhaoming Lu \corref{mycorrespondingauthor}}
\cortext[mycorrespondingauthor]{Corresponding author}
\ead{lzy0372@bupt.edu.cn}
\author[mymainaddress,mysecondaryaddress,mythirdaddress]{ Shuang Zhou }
\ead{zhoushuang@bupt.edu.cn}
\author[mymainaddress,mysecondaryaddress,mythirdaddress]{\\ Xiangming Wen }
\ead{xiangmw@bupt.edu.cn}
\author[mymainaddress,mysecondaryaddress,mythirdaddress]{ Zhihong He }
\ead{hezhihong@bupt.edu.cn}
\address[mymainaddress]{Beijing Laboratory of Advanced Information Networks}
\address[mysecondaryaddress]{Beijing Key Laboratory of Network System Architecture and Convergence}
\address[mythirdaddress]{School of Information and Communication Engineering,
Beijing University of Posts and Telecommunications, Beijing, China.}

\begin{abstract}
  As elderly population grows, social and health care begin to face validation challenges, in-home monitoring is becoming a focus for professionals in the field. Governments urgently need to improve the quality of healthcare services at lower costs while ensuring the comfort and independence of the elderly. This work presents an in-home monitoring approach based on off-the-shelf WiFi, which is low-costs, non-wearable and makes all-round daily healthcare information available to caregivers. The proposed approach can capture fine-grained human pose figures even through a wall and track detailed respiration status simultaneously by off-the-shelf WiFi devices. Based on them, behavioral data, physiological data and the derived information (e.g., abnormal events and underlying diseases), of the elderly could be seen by caregivers directly. We design a series of signal processing methods and a neural network to capture human pose figures and extract respiration status curves from WiFi Channel State Information (CSI). Extensive experiments are conducted and according to the results, off-the-shelf WiFi devices are capable of capturing fine-grained human pose figures, similar to cameras, even through a wall and track accurate respiration status, thus demonstrating the effectiveness and feasibility of our approach for in-home monitoring.
\end{abstract}

\begin{keyword}
In-home monitoring \sep Healthcare \sep WiFi \sep CSI \sep Capturing human pose figures \sep Tracking respiration status
%\texttt{elsarticle.cls}\sep \LaTeX\sep Elsevier \sep template
%\MSC[2010] 00-01\sep  99-00
\end{keyword}

\end{frontmatter}

% \linenumbers

\section{Introduction}\label{sec1}

With the improving of the medical technical level, the global population starts to take on the growth of the outbreak, while the demand for elderly care is rising. According to the statistical results, by 2030, the percentage of the population over 60 will double, i.e., about 20\% of the total population \cite{fowles1986profile}. As a result, the elderly population is becoming a worldwide concern rapidly. With the rising cost of medical care, the current quality of services is obviously unable to meet the needs of The Times. In-home monitoring for healthcare provides a possible solution to overcome these challenges.

On the other hand, according to a survey conducted in the United States, about 92\% of the elderly live alone, with about 50\% over the age of 75 \cite{fowles1986profile}. While surveys undertaken in China show that 50\% of the households in China have elders who live alone \cite{agingdata}. Furthermore, surveys undertaken in European Union reveal that by 2050, the dependency ratio of the elderly --- the ratio of people over 64 to those aged 15 and 64 --- is expected to double. This
suggests that the old days of raising two elders per working-age person in the European Union are over. The statistical results clearly demonstrate the necessity and urgency of in-home monitoring needs, which can make it easier for caregivers to access daily healthcare information and enables the elderly to live independently.

Existing in-home monitoring systems are accomplished by 1) record physiological data with wearable devices and wearable sensors, and use these devices as dynamic monitors; 2) embed sensors (e.g., cameras) for behavioral data collection into furniture and home environment; 3) integration of the above two methods \cite{tamura1988bed}. Obviously, part of the systems is limited since they can only obtain physiological data or behavioral data. And more importantly, these kinds of systems are always made up of sensors of different functions or types, wireless actuators, software applications, computer hardware, computer networks and smart phones. These devices are mainly connected to each other to exchange data and provide services. Therefore, it is expensive to deploy these systems in practical. In addition, camera-based systems raise privacy concerns and are susceptible to lighting conditions and obstacles like walls, while wearable sensors are inconvenient for the users.

The reason of the study is the main source for the elderly daily healthcare related working interest, in other words, our core purpose is to develop a new, convenient and cheap in-home monitoring approach, which can make medical services available to more elderly people, and to promote independence and comfort of the elderly as much as possible at the same time.
\begin{figure*}[htbp]
\centerline{\noindent\makebox[\textwidth] {
    \includegraphics[width=0.63\paperwidth]{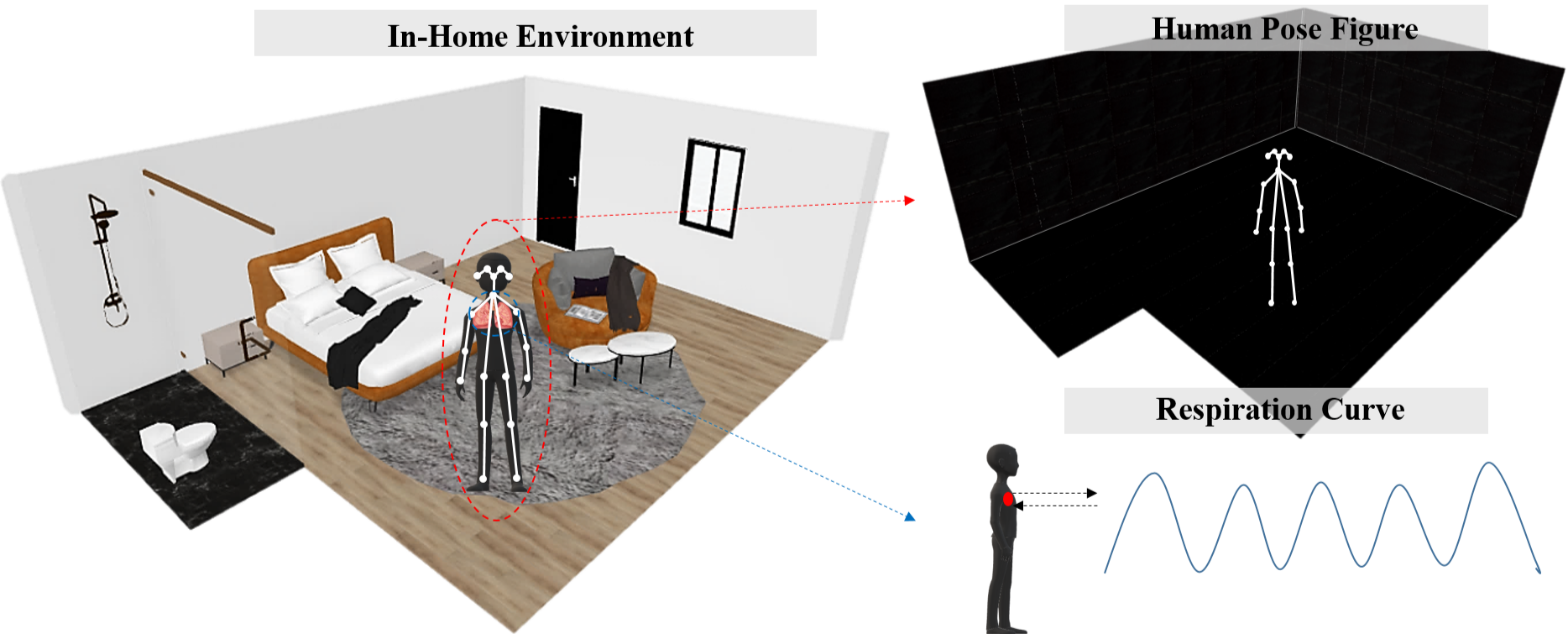}}}
\caption{\scriptsize{Sketches of the approach output. The left part shows a typical indoor environment, while the right part demonstrates sketches of a human pose figure captured and a respiration curve tracked by the proposed approach.}}

\label{fig111}
\end{figure*}

So we are proposing a low-costs, non-wearable and privacy-protection approach which enables all-round monitoring for the daily healthcare of the elderly. Our approach mainly realizes the capture of fine-grained human pose figures (behavioral data) even through a wall, and track the vital signs (physiological data), especially detailed respiration status of humans by several ubiquitous off-the-shelf WiFi devices. Capturing human pose figures means that we can produce a fine-grained and accurate description of human poses by off-the-shelf WiFi devices, similar to that achieved by cameras. While tracking respiration status means that we can use off-the-shelf WiFi devices to track the heaving chest and produce a respiration status curve. Figure \ref{fig111} shows the sketches of a human pose figure and a respiration status curve outputted by the designed approach. According to the human pose figures captured, existing computer vision approaches can be used to monitor the elderly, i.e., tracking and detecting abnormal events like remaining motionless, fall-down, stagger, panic gesturing, moving around more slowly, etc. \cite{MiSuen2003Computer}. Abnormal respiration patterns and underlying diseases can also be detected and distinguished according to the respiration status curves tracked to provide medical support for pre-diagnosis \cite{forkan2016probabilistic}.
\begin{figure*}[htbp]
\centerline{\noindent\makebox[\textwidth] {
    \includegraphics[width=0.63\paperwidth]{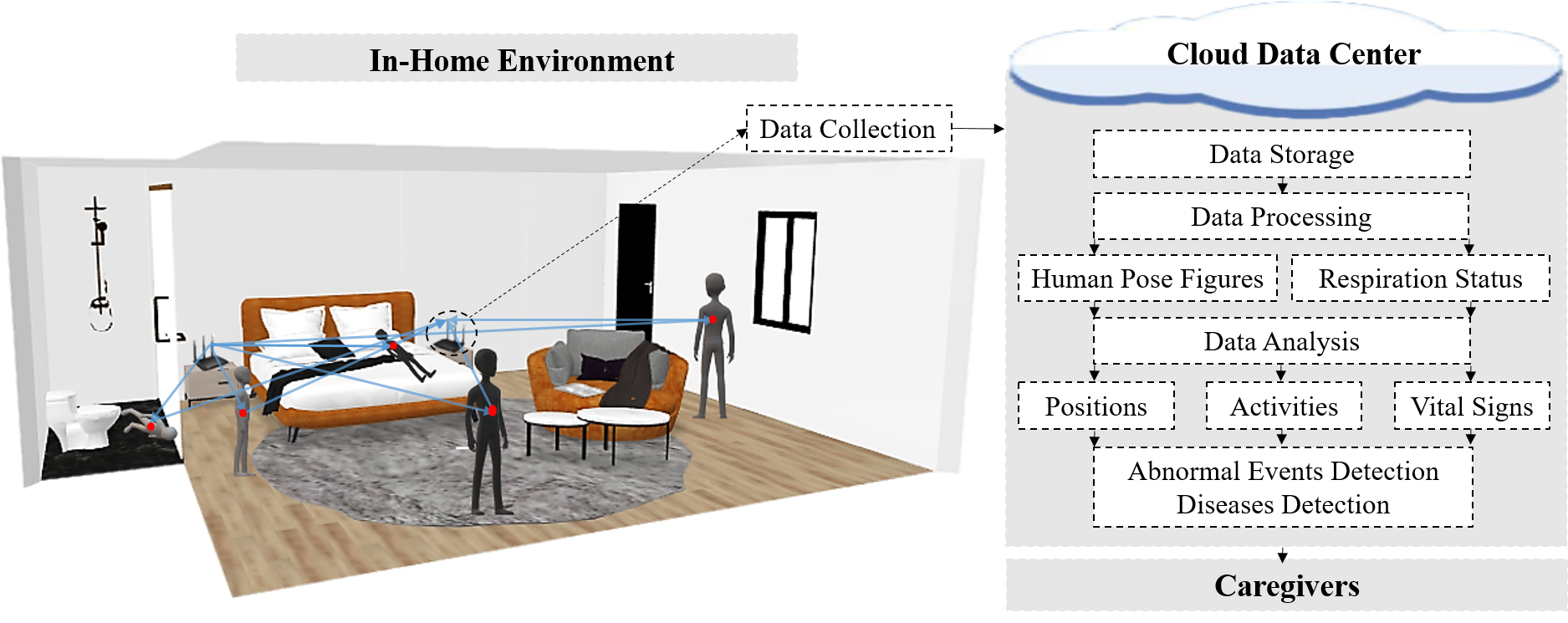}}}
\caption{\scriptsize{Overview of the proposed approach. The left part shows how WiFi signals propagate in a typical indoor environment and how to collect CSI data, while the right part demonstrates a brief flow of the approach.}}
\label{intro}
\end{figure*}

Leveraging the prevalence of WiFi infrastructures and Channel State Information (CSI) obtained from off-the-shelf WiFi cards \cite{intel5300}, we design the elderly-focused in-home monitoring approach, as shown in Figure \ref{intro}. Specifically, CSI mainly represents the multipaths in which WiFi signals are propagated, reflected, diffracted and scattered in a typical indoor environment. Thus it can capture how WiFi signals travel through humans. The receivers collect CSI and send it to the cloud data center, where the CSI is stored, processed and analyzed. In the cloud, fine-grained human pose figures and detailed respiration status are extracted, and with which more information about daily healthcare can be obtained, such as positions, activities and vital signs, etc. Then abnormal events and underlying diseases could be detected consequently. Note that all the above information is available according to the demand by caregivers, thus caregivers can see all-round daily healthcare information about the elderly easily and directly.

Existing off-the-shelf WiFi-based human sensing systems provide only a snapshot of information about humans. To be Specific, some systems simplify the human into a single point reflector for tracking or localization, while others classify or recognize limited activities by matching pre-training samples \cite{tan2019multitrack}. And past breathing detection systems focus on the respiratory rate over a period rather \cite{farsense} than the time domain respiration status curves, which are quite limited when utilized to timely detect abnormal respiration and underlying diseases. Extracting detailed, accurate information about humans, such as pose figures and respiration status curves, is much more challenging:
\begin{itemize}
\item The bandwidth and number of antennas of off-the-shelf WiFi devices are limited, which can not provide adequate resolution \cite{WiRIM} for capturing fine-grained human pose figures and detailed respiration status curves.
\item Since WiFi signals are not visible, it is impossible to annotate human poses manually, which makes the mature approaches based on computer vision \cite{alp2018densepose,fang2017rmpe} not suitable for capturing human pose figures by WiFi.
\item In an indoor environment, the received WiFi signals is an aggregation of all the multipaths \cite{zhang2019breathtrack}. Therefore, we need to extract valid and accurate signal components related to human poses or human respiration.
\end{itemize}

In view of the above challenges, we propose a corresponding multi-device cooperative working system, and fully use both CSI amplitude information and CSI phase information to improve the resolution of off-the-shelf WiFi devices. In order to capture fine-grained figures of human poses, we firstly extract valid and accurate CSI related to human poses, based on which we construct CSI maps. Meanwhile, a synchronized camera is used to extract human skeletons and annotate WiFi signals. A neural network is designed to map the CSI maps to human pose figures. For tracking detailed respiration status, we design a series of signal processing algorithms to obtain the respiration status curves and detect the abnormal respiration period. We build a prototype system for performance experiments. According to the results of the experiments, off-the-shelf WiFi devices are capable of capturing human poses with the same granularity as cameras even through a wall and can achieve 80\% correlation with respiration ground truth for respiration status curve tracking. The main contributions of this work include the following aspects:
\begin{itemize}
  \item  We provide a new approach for in-home monitoring which makes all-round daily healthcare information (both behavioral information and physiological information) of the elderly available to caregivers using only off-the-shelf WiFi devices. Compared with other in-home monitoring systems, the proposed approach is low-costs, non-wearable and privacy-protection.
  %We provide a low-costs, non-wearable and privacy-protection approach for in-home monitoring, which makes all-round healthcare information (both behavioral information and physiological information) about the elderly available to caregivers using only off-the-shelf WiFi devices.
  %\item The proposed approach can capture fine-grained human pose figures and track respiration status curves simultaneously based on off-the-shelf WiFi.
  \item As the basis for behavioral information, the proposed approach is capable of capturing fine-grained human pose figures even through a wall, and the accuracy is comparable to that achieved by cameras. To the best of our knowledge, this is the first work to capture fine-grained human pose figures through a wall by off-the-shelf WiFi devices.
  \item As the basis for physiological information, the proposed approach can track detailed respiration status curves. It can detect abnormal respiration (apnea) using the respiration status curves, and can classify abnormal respiration patterns include apnea accordingly, providing medical support for pre-diagnosis.
\end{itemize}

The rest of this work consists of the following parts. Related work and preliminaries are discussed in Section \ref{sec2} and Section \ref{sec3}, respectively. Section \ref{sec4} provides an overview to the approach, followed by Section \ref{sec5} and Section \ref{sec6} describe the two most important parts of the approach, i.e., capturing human pose figures and tracking detailed respiration status, respectively. Implementation details are presented in Section \ref{sec7}. Section \ref{sec8} describes the performance of our approach. Section \ref{sec9} proposes some discussions and Section \ref{sec10} concludes this work.

%\paragraph{Installation} If the document class \emph{elsarticle} is not available on your computer, you can download and install the system package \emph{texlive-publishers} (Linux) or install the \LaTeX\ package \emph{elsarticle} using the package manager of your \TeX\ installation, which is typically \TeX\ Live or Mik\TeX.
%\paragraph{Usage} Once the package is properly installed, you can use the document class \emph{elsarticle} to create a manuscript. Please make sure that your manuscript follows the guidelines in the Guide for Authors of the relevant journal. It is not necessary to typeset your manuscript in exactly the same way as an article, unless you are submitting to a camera-ready copy (CRC) journal.
%\paragraph{Functionality} The Elsevier article class is based on the standard article class and supports almost all of the functionality of that class. In addition, it features commands and options to format the
%\begin{itemize}
%\item document style
%\item baselineskip
%\item front matter
%\item keywords and MSC codes
%\item theorems, definitions and proofs
%\item lables of enumerations
%\item citation style and labeling.
%\end{itemize}
\section{Related Work}\label{sec2}
In this section, we firstly introduce the representative in-home monitoring systems to further illustrate the motivation and significance of our work. Then we discuss some relevant sensing work using off-the-shelf WiFi and some methods for capturing human pose figures to show the technical background of our work.

\subsection{In-Home Monitoring}
With the further development of the most relevant research technology, the research field of in-home monitoring systems applied to the elderly living alone has been expanded. A professional functional floor, ultrasonic sensors, videos, robots and other types of sensors are used to monitor the life of the elderly in the project called Aware Home \cite{alsina2017homesound,cook2004smart}. There are many other similar projects such as Ubiquitous Home \cite{yamazaki2007ubiquitous} that aim to provide a comprehensive in-home monitoring of the elderly. However, the installation of such systems is not only disruptive, but also costly, making the systems inaccessible to many elders.

The EasyLiving project \cite{brumitt2000easyliving} in Microsoft Research uses a computer vision approach to check the positions or identities of humans in a room. It can be used in applications that aid everyday tasks in indoor spaces, which is much cheaper, but raises privacy concerns. CoSHIE \cite{pham2016cloud} collects physiological data through wearable sensors, and WonedRing \cite{zhou2015healthcare} is a kind of wearable sensor used for analysis and detection of user behaviors. However, wearable sensors are not suitable and convenient for elderly healthcare. Worse, the elderly may forget to put on the wearable devices once they have taken them off. What's more, these systems can only obtain physiological or behavioral data, which makes it difficult for them to provide comprehensive monitoring for the elderly.

Thus, a low-costs, non-wearable and privacy-protection in-home monitoring solution for all-round healthcare of the elderly is urgently needed.

\subsection{Human Sensing with Off-the-Shelf WiFi}
Here, we mainly study and introduce some representative human sensing applications based on off-the-shelf WiFi, and these applications generally include the following three categories.

\paragraph{Activity Recognition} CARM \cite{CARM} focuses on the creation of a CSI activity model used to recognize human activities. WiDance \cite{qian2017inferring} creatively captures complete information corresponding to the Doppler frequency shifts caused by human movements, and creates a prototype of contactless dance game. WiFall \cite{wang2016wifall} completes high-precision level fall detection. Wi-Chase \cite{arshad2017wi} can carry out the applicable subcarriers in the WiFi signals and uses them in the recognition of human activities. WiFit \cite{li2018wifit} recognizes the exercise types, and is able to calculate the sporting quantity of different groups under different environmental conditions. \cite{chen2018wifi} realizes human activity recognition on the basis of an attention-based Bidirectional Long Short-Term Memory (Bi-LSTM) network, and reaches the highest recognition accuracy of different activities compared to other methods. \cite{xu2019indoor} uses the temporal information contained in the CSI time series to monitor events in different indoor environments. WiFiMap+ \cite{zhang2019wifimap+} recognizes high-level indoor semantics in the environments and human activities based on WiFi signals.

However, these systems can only recognize a few predefined activities in fixed positions, the robustness and practicability are not enough.

\paragraph{Localization or Tracking} Widar \cite{qian2016decimeter}, \cite{qian2017widar} mainly uses CSI dynamics to conduct human speed tracking and human localization. Widar2.0 \cite{qian2018widar2} develops an efficient algorithm and uses it to estimate Doppler frequency shifts, Angle of Arrival (AoA), Time of Flight (ToF) and other parameters. At the same time, the original parameters are converted into a high accuracy position through a designed pipeline. IndoTrack \cite{indotrack} and \cite{li2018training} use AoA and spatial temporal Doppler frequency shifts for accurate human tracking. PADS \cite{qian2018enabling} leverages spatial diversity across multiple antennas and all CSI information (including phase and amplitude) to adjust and extract sensitive indicators, and finally realizes not only robust but also accurate target detection.

These systems abstract a human into a single point reflector so as to realize the localizing, tracking, and even monitoring the walking speed of the human body.

All in all, in the past, off-the-shelf WiFi-based sensing systems could only be limited to capturing part of a human's information. Therefore, this work mainly attempts to use off-the-shelf WiFi devices like cameras, which can directly capture fine-grained figures of human poses.

\paragraph{Respiration Detection} Wi-Sleep \cite{Wi-Sleep} is the first system that utilizes CSI amplitude for sleep breathing detection and its subsequent work \cite{Contactless} adds sleep posture and sleep apnea detection module. Phasebeat \cite{Phasebeat} mainly uses the CSI phase difference between two receiving antennas to capture respiration. The main concern of these systems is the difference in human respiration rates at a given period of time, not the detailed breath status. \cite{FZmodel} mainly introduces the Fresnel Zone model, based on which a respiration sensing model using WiFi is constructed. According to the Fresnel zone model, respiration detection based on CSI amplitude may fail in some areas. FullBreathe \cite{Fullbreathe} aims to address the undetectable region problem by exploiting the complementary property between CSI amplitude and phase data, but it presents the detection ability ratio metric instead of detailed respiration status to evaluate system performance. Farsense \cite{farsense} employs the ratio of CSI from two antennas and also leverages the complementary property between CSI amplitude and phase to eliminate the ``blind spots” problem and expands the sensing range, but it focuses on sense range rather than detailed respiration status. BreathTrack \cite{zhang2019breathtrack} tracks the detailed respiration status, but it utilizes a hardware correction method to obtain accurate CSI, which limits its usage in real life.

For in-home monitoring, respiration status need to be tracked as detailed as possible, while the equipment should be as simple as possible. Therefore, in this work, we try to track detailed respiration status curves and detect abnormal respiration with only off-the-shelf WiFi devices.

\subsection{Capturing Human Pose figures}
Since capturing human pose figures by off-the-shelf WiFi is very pioneering, we have little work to refer. Here we introduce some representative work, which is naturally divided into the following two categories.

\paragraph{Computer Vision} Capturing human poses from images is a known problem called human pose estimation in the computer vision literature, such as DensePose \cite{alp2018densepose}, AlphaPose \cite{fang2017rmpe} and CPN \cite{chen2018cascaded}, which infers the human position from an image then regressing the keypoint heatmaps.

But the camera-based approaches are sensitive to light conditions as well as obstacles, and raise privacy concerns if used at home.

\paragraph{Wireless Systems}
Recently, researchers have paid more attention to estimate human poses using wireless signals. RF-Pose \cite{zhao2018through} utilizes a radar implemented with frequency modulated continuous wave (FMCW) equipment \cite{adib2015capturing} to estimate human poses, and so does RF-Pose3D \cite{zhao2018rf}. The equipment works in WiFi frequencies (5.46-7.24$GHz$) and each antenna array of it utilizes 4 transmitting antennas and 16 receiving antennas to improve the spatial resolution.
All these are not available on off-the-shelf WiFi devices, which makes it difficult to estimate human poses by off-the-shelf WiFi devices (see Section \ref{sec3} for details). Besides, the equipment is expensive and unavailable to the elderly.
%Therefore, the equipment is not available in daily life and much more expensive than off-the-shelf WiFi devices.

On the other hand, our previous work \cite{8943379} %and \cite{wang2019can}
illustrates the possibility of using WiFi devices for human pose estimation, but
%the performance is limited and they
it only verifies the performance in visible scenarios. For in-home monitoring, this work tries to make off-the-shelf WiFi devices capture figures of human poses with higher accuracy even through a wall.

\section{Preliminaries}\label{sec3}
\subsection{WiFi Signals Acquisition and Properties}
\paragraph{Acquisition}
Our in-home monitoring approach relies on transmitting WiFi signals and receiving its reflections by off-the-shelf WiFi devices. CSI is widely used to describe the transmission of WiFi signals between a pair of transmitter and receiver, which refers to the multipath propagation of some carrier frequencies \cite{ma2019wifi}. CSI measurements can be obtained from the received packets based on the Intel 5300 NIC with modified firmware and driver\cite{intel5300}.

\paragraph{Properties}
%Besides, WiFi signals have some intrinsic features, which makes our work more challenging and worthy.
%\begin{itemize}
  %\item
  Compared to other wireless devices used for capturing human poses and tracking respiration  \cite{zhao2018through}, \cite{zhao2018rf}, \cite{extracting2018}, which work in 5.46-7.24\(GHz\), the bandwidth of 802.11n (20\(MHz\) or 40\(MHz\)) is insufficient \cite{xie2015precise} to capture human poses accurately and track detailed respiration status. At the same time, for off-the-shelf WiFi cards, the number of antennas is limited, which results in a limitation of spatial resolution.
  %\item
  Besides, WiFi signals frequency range can traverse obstacles like walls, etc., but in the frequency range of WiFi signals, the human is a kind of specular reflector \cite{scattering1987}. Consequently, our approach can be used in through-wall scenarios.
%\end{itemize}

\subsection{Channel State Information}
\paragraph{Overview of CSI} CSI represents the samples of Channel Frequency Response (CFR) in each Orthogonal Frequency Division Multiplexing (OFDM) subcarrier:
\begin{equation}
\label{CSIoverview}
    H(f,t)=\sum^K_{i=1} H_i(f,t)=\sum^K_{i=1} \alpha_i(t)e^{-j2\pi f\tau_i(t)},
\end{equation}
where $K$ means the number of paths and $H_i(f,t)$ is the channel response of the $i$-th path at time $t$. $\alpha(t)$ is the attenuation and $\tau(t)$ means the propagation delay.

To be specific, CSI is a three-dimensional matrix of complex values. One CSI measurement specifies the amplitude and phase of the channel response for the corresponding subcarrier between a single transmitter-receiver antenna pair. Furthermore, $N$ CSI is measured for all the 30 subcarriers, and a complex vector is finally formed:
\begin{equation}
\label{CSIsample}
    H(i)=[H_1(i), H_2(i),\dots,H_{30}(i)], \quad i=1,2,\dots,N
\end{equation}
where $H_k(i)$ contains the amplitude and phase information.

A time serie CSI measurements can capture how wireless signals travel through surrounding humans and objects in the space domain, time domain and frequency domain. Therefore, it can be  applied in different wireless sensing systems \cite{ma2019wifi}. For example, as the amplitudes of CSI vary in the time domain resulting in different patterns for different postures or gestures, they can be applied to recognize postures or gestures. Signal transmission direction and delay are corresponding to the phase shifts of CSI, which can be used for human localization and tracking.

\paragraph{Phase Offset Removal} According to the description in \cite{Fullbreathe}, CSI can be divided into static and dynamic components. Among them, the static component $H_s(f,t)$ mainly consists of the Line of Sight (LoS) path and other reflection paths from static objects, while the dynamic component $H_d(f,t)$ covers the paths reflected from the moving body parts or a human's chest who remains still. The dynamic component can be sheltered by the static component since the frequency response of LoS path is much stronger than other reflection paths. Make $G_d$ represent the set of dynamic paths, so Equation \ref{CSIoverview} can be rewritten as follows:
\begin{equation}
\label{S&D path}
    H(f,t)=H_s(f,t)+H_d(f,t)
    =H_s(f,t)+\sum_{i\in G_d} \alpha_i(t) e^{-j2\pi f\tau_i(t)}.
\end{equation}

Due to hardware imperfection of off-the-shelf WiFi devices, different time-varying phase offsets are often included in consecutive CSI measurements \cite{atheros}. In this work, we use Conjugate Multiplication (CM) of CSI between antennas to eliminate the phase offset\cite{indotrack}:
\begin{equation}
\label{CM}
    H_{CM}(f,t)=H_1(f,t)\bar{H}_2(f,t) \\
	=H_{1,s}\bar{H}_{2,s}+H_{1,d}\bar{H}_{2,d}+H_{1,s}\bar{H}_{2,d}+H_{1,d}\bar{H}_{2,s},
\end{equation}
where $H_1(f,t)$ refers to the CSI on the first antenna and $\bar{H_2}(f,t)$ refers to the conjugate CSI on the second antenna. By referring to Equation \ref{S&D path} and Equation \ref{CM}, we can know that $H_{1,s}\bar{H}_{2,s}$ represents the product of corresponding static components between two different antennas, so we can treat it as a constant if the time is short enough. $H_{1,d}\bar{H}_{2,d}$ refers to the product of corresponding dynamic components on the antennas. Comparing with others, this term is relatively small, so it can be ignored. The rest two terms are what we need to focus on.

\paragraph{Fresnel Zone Model}
CSI amplitude and phase are not only affected by one path, but multipaths. According to the Fresnel zone model, a pair of transmitter and receiver and the surrounding space are divided into concentric ellipses, which are called Fresnel zone regions. Fresnel zone model reveals the propagation and the deflection of WiFi signals in the Fresnel zone regions. At the same time, different path lengths result in different amplitude attenuation and phase shift, which leads to the constructive and destructive effect at the receiver.

If an object moves in multiple Fresnel zone regions, the signal displayed in the receiver will take on the form of a sine wave. In addition, it is considered that the best location for CSI amplitude-based respiration sensing is in the middle of a Fresnel zone region, while the worst is at the boundary \cite{FZmodel}. Reference \cite{Fullbreathe} theoretically and experimentally shows that CSI amplitude and CSI phase are orthogonal and complementary to each other. CSI phase changes significantly at the boundary of Fresnel zone regions, but merely changes in the middle. This phenomenon suggests that a bad (good) position for respiration sensing in amplitude is a good (bad) one in phase.

\subsection{Convolutional Neural Network}
\paragraph{Deep Neural Network} There are many layers of neurons in a deep neural network, and input information can be extracted through them. Input information can be obtained from the previous layer for each neural. And nonlinear functions and weights are used to combine the obtained information.

\paragraph{Convolutional Neural Network} For a Convolutional Neural Network (CNN), each neuron contained in it is related to several neurons in the previous layer. And this is the significant difference between CNNs and the ordinary neural networks. For CNNs, all the neurons in the same layer are equally weighted. So the computation of the neuron values can be thought of as the convolution of a weight kernel and the neurons from the previous layer. CNNs make sure the local independence of data to reduce the computational complexity, which make deeper networks possible accordingly.

\paragraph{Resize Convolution} When generating images, neural networks are typically built against high levels of description and low resolution and then fill in the details. The so-called deconvolution operation refers to the method of converting low-resolution images to obtain high-resolution images. However, for deconvolution, there is often uneven overlap, which is generally referred to as ``Checkerboard Artifacts" \cite{odena2016deconvolution}. One approach to solve this problem is basically to resize the image and then do a convolution. This approach is called resize convolution, a roughly similar method works well in image super-resolution \cite{aitken2017checkerboard}.

\paragraph{Squeeze-and-Excitation Block} The convolution operator of CNNs uses spatial information and channel information in the local receptive fields of each layer to enable the network to construct information features. The main purpose of Squeeze-and-Excitation (SE) block is to improve the quality of the representations extracted by a neural network by modelling the interdependencies between the convolution feature channels. It mainly emphasizes the useful information and suppresses the less useful ones by performing feature recalibration \cite{senet2018}.

\begin{figure}[!htpb]
\centerline{\noindent\makebox[\textwidth] {
    \includegraphics[width=0.63\paperwidth]{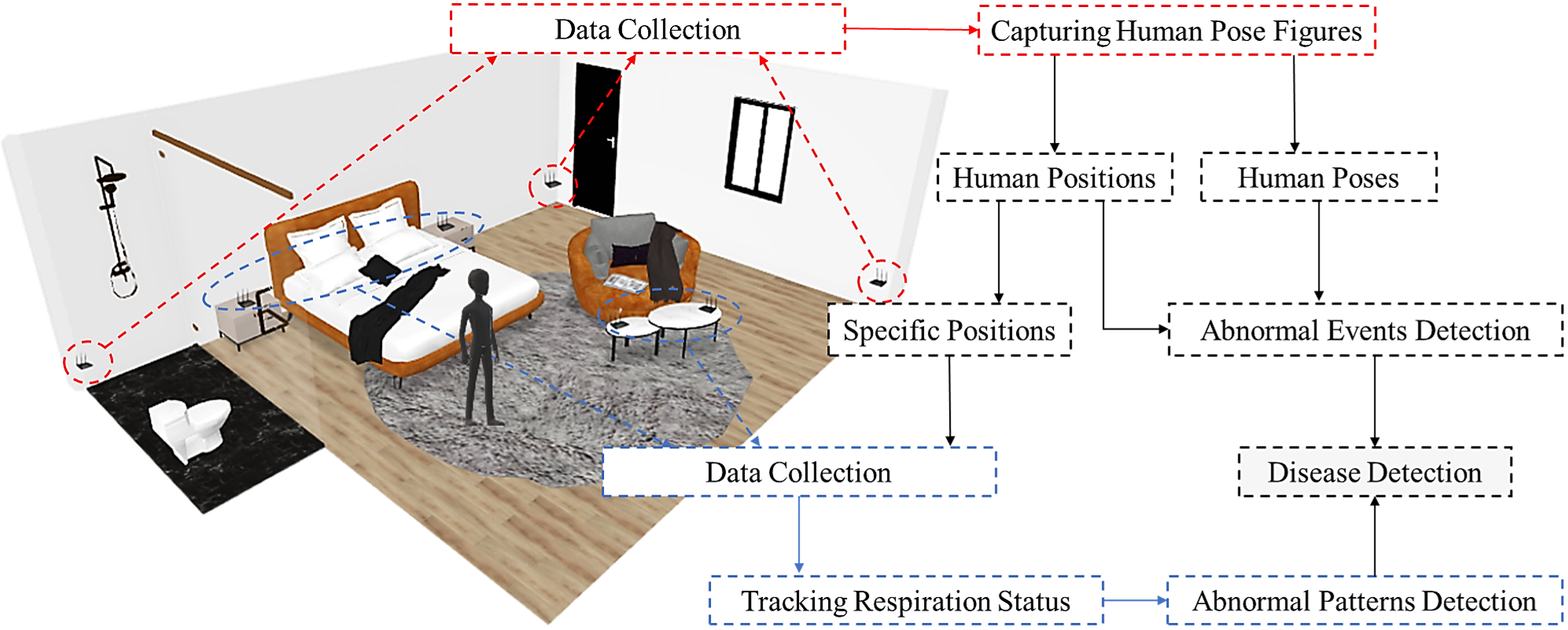}}}
\caption{\scriptsize{Architecture of the proposed approach. The left part shows a typical indoor environment, while the right part demonstrates the approach flow in detail.}}
\label{fig1}
\end{figure}
\section{Approach Overview}\label{sec4}
Figure \ref{fig1} shows the details of the approach in this work. According to \cite{wu2017device}, when the LoS path between a pair of transceivers and the walk path of a human are parallel, the transceivers will be unable to realize sensing the human. Consequently, this work collects data through three off-the-shelf WiFi devices (circled in red) and adopts the following method to improve the spatial resolution of off-the-shelf WiFi: the two pairs of transceivers are placed perpendicular to each other to provide information from all perspectives for accurately capturing human pose figures. At the same time, we need to take full advantage of the CSI amplitude and phase, since human pose information is mainly included in CSI amplitude, and human position information is mainly included in CSI phase. Capturing human pose figures can provide not only human pose information but also human position information, which can be used to detect abnormal events.

And if the target human appears in specific positions, i.e., on the bed, on the sofa, in the bathroom or other dangerous places, the devices (circled in blue) there will start to collect data for respiration tracking. We can obtain detailed respiratory status curves, with which abnormal respiration patterns can be detected. Similarly, we also use both CSI amplitude and CSI phase, because CSI amplitude and CSI phase are orthogonal and complementary to each other.

Due to the mature development of human pose and position recognition technology and abnormal event detection technology based on computer vision, this work will mainly demonstrate how to capture human poses by off-the-shelf WiFi as well as how to track respiration status and detect abnormal patterns by off-the-shelf WiFi. The deployment of the whole approach will be explained briefly in Section \ref{sec9} and Section \ref{sec10}.

\section{Capturing Human Pose Figures}\label{sec5}
WiFi CSI has no direct information about human poses. And it is impossible to annotate human poses in WiFi CSI manually. Therefore, a neural network is designed to solve this problem, which converts WiFi signals into figures by synchronizing a camera to extract human skeletons and annotating WiFi CSI during the training process. After completing the training, the system is capable of capturing human pose figures using only WiFi CSI as input. In addition, because WiFi signals can traverse obstacles, the system we build can also capture human pose figures even through a wall.

Beyond neural network design, the design of capturing human pose figures accounts for intrinsic features of CSI, which are listed as follows.
\begin{itemize}
  \item As mentioned in Section \ref{sec3}, off-the-shelf WiFi devices tend to have a low spatial resolution.
  \item Since the pose features involved in the dynamic component are significantly weaker than the static component, and are sometimes interfered by irrelevant noises, CSI cannot be directly used to capture human poses.
  \item The representations and perspectives of CSI and supervisory video frames are different.
\end{itemize}

Therefore, we leverage three off-the-shelf WiFi devices and the CSI phase and amplitude together in order to improve the resolution. And we design signal processing methods, whose main purpose is to extract valid and accurate CSI related to human poses and use it in the construction of CSI maps firstly and then use the constructed CSI maps as input of the neural network. We will mainly cover signal processing and neural network design in detail separately in this section.

\subsection{Signal Processing}
\paragraph{Antenna Selection} It is observed that compared to other antennas, there is always an antenna with larger dynamic responses. While there is also an antenna which is insensitive, mainly composed of static component. Inspired by the observation, the most sensitive antenna is regarded as the reference receiving antenna, and the data of the least sensitive antenna is discarded. Note that we judge the sensitivity of the antenna according to the variance value.

\begin{figure*}[htpb]
\centerline{\includegraphics[width=0.64\paperwidth]{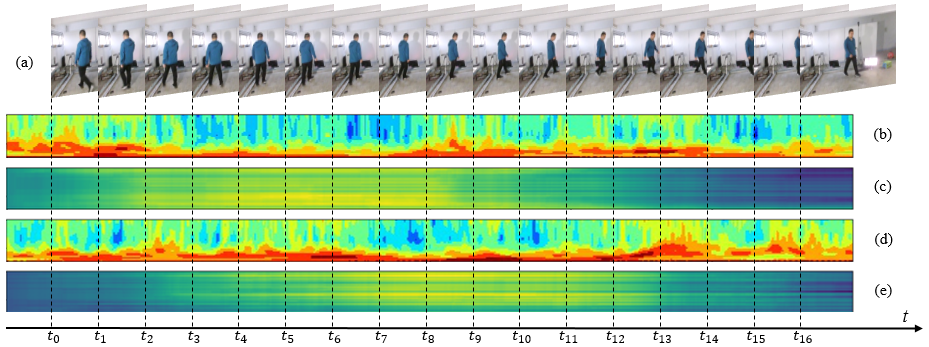}}
\caption{\scriptsize{Examples of video frames and corresponding CSI maps over time. (a): Video frames of a volunteer over time. (b) (d): Temporal-frequency features extracted from CSI amplitude at different receivers. (c) (e): Relative phases at different receivers. Note that the vertical axises in (b)-(e) are the subcarrier indices ranging from 0-29. If the pixel value is higher, then the corresponding map will be darker.}}
\label{4bar}
\end{figure*}
\paragraph{CSI Map Construction} As you may recall, we focus on constructing CSI maps with both CSI amplitude and CSI phase. For one thing, Principal Component Analysis (PCA) is firstly applied on the CSI amplitude of the two remaining antennas to remove redundant and unrelated information, and the key human pose information will be retained. At the same time, the second principal component will be carried out, considering it can clearly capture the changes of human poses. There is a correlation between the specific frequency and the rate of length changes of the reflection paths corresponding to humans \cite{CARM}, so we mainly utilize Discrete Wavelet Transform (DWT) to extract the temporal-frequency features contained in the second principal component. The temporal-frequency features extracted from CSI amplitude are shown in Figure~\ref{4bar} ((b) and (d)).

For another, as mentioned in Section \ref{sec3}, conjugate multiplication method is mainly used to deal with the time-variant random phase offset on the two retained antennas in our work. We find that only the random phase offset can be eliminated by the application of conjugate multiplication method, while we mainly aim at extracting the dynamic component related to human poses directly. It should be pointed out that in Equation \ref{CM}, if the time is short enough, $H_{1,s}\bar{H}_{2,s}$ can be treated as a constant and can be eliminated by subtracting the average. The residual terms are two products of the dynamic component of one antenna and the static component of another antenna.
We mainly increase the static component power of the first antenna by adding a value \(\delta \) and reduce the static component power of the second antenna by subtracting \(\gamma \). To be specific, the main parameter we need to determine is \(\gamma \), we set it so as to ensure the minimum amplitude of all the CSI samples corresponding to the second antenna is $0$. And we set
\(\delta {\rm{ = }}1000\gamma \). It should be pointed out that the reference antenna is set as the second antenna, while the rest antenna is set as the first. Figure~\ref{4bar} ((c) and (e)) illustrates the relative CSI phases over all the subcarriers after processing.

Moreover, as shown in Figure~\ref{4bar}, the temporal-frequency features extracted from CSI amplitude and relative CSI phases over time are significant different, that is why CSI can be used for capturing human pose figures.

\begin{figure}[htpb]
\centerline{\noindent\makebox[\textwidth] {
    \includegraphics[width=0.42\paperwidth]{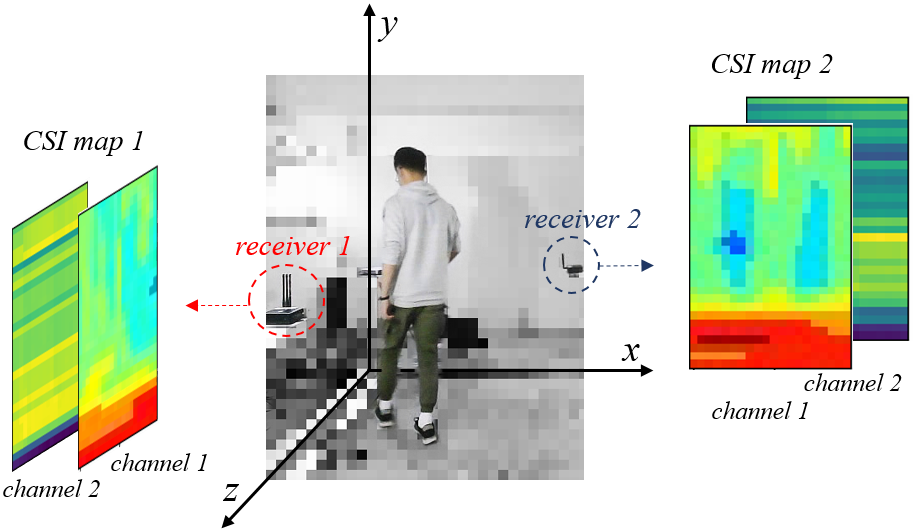}}}
\caption{\scriptsize{Constructed CSI maps from the two receivers (two perspectives) and the corresponding video frame recorded by a camera at the same time. Channel 1 is the amplitude channel, while Channel 2 is the relative phase channel.}}
\label{3d}
\end{figure}
Then we construct a novel sensitive CSI map, which is mainly composed of an amplitude channel and a relative phase channel. \(M \times T\) pixels are contained in each channel, where \(M\) and \(T\) respectively represent the number of subcarriers and the length corresponding to a specific time segment. Recall that the spatial resolution of off-the-shelf WiFi is low, thus we utilize multiple CSI samples for CSI map construction. By doing so, we can not only make the captured human poses more complete, but also depict the correlation between continuous human poses more precisely. We use 20 CSI samples for one CSI map in this work, which means \(T = 20\). Figure~\ref{3d} shows the CSI maps which are constructed from the two receivers (two perspectives) and the corresponding video frame recorded by a camera at the same time. We will explain \(T\) in detail in Section \ref{sec7}.

\subsection{Neural Network Design}
\paragraph{Data and Annotations} Considering that it is impossible to manually annotate human poses in CSI maps, thus we capture synchronized videos through a camera connected to one of the receivers. To this end, we mainly use OpenPose \cite{cao2018openpose} to extract the human skeletons from the videos, so as to provide ground truth annotations for CSI. It is important to note that our primary goal is to get the skeletons, not the keypoints, as skeletons are more fault tolerant. This is because the misjudgment of one pixel has little effect on the whole human pose figure while the misjudgment of a keypoint will lead to an incorrect human pose. And because of the limited spatial resolution of off-the-shelf WiFi, using skeletons as annotations will be more accurate than using keypoints as annotations. Moreover, we use greyscale human skeleton figures as annotations to simplify the design of the neural network.

\paragraph{Network Framework}
We need to take human spatial positions and temporal correlation of human poses into consideration in the design process of the neural network. In addition, considering the limited spatial resolution of off-the-shelf WiFi, the network needs to learn the aggregate information contained in multiple CSI samples, as mentioned earlier. Thus CNNs are basically utilized to extract the time and space features.

For the network, it is necessary to convert the information from the view of the off-the-shelf WiFi devices into the information from the view of the camera. Therefore, our network firstly obtains features from the CSI samples, and then decodes the features through the view of the camera to obtain human skeletons.

\begin{figure*}[!t]
\centerline{\includegraphics[width=0.64\paperwidth]{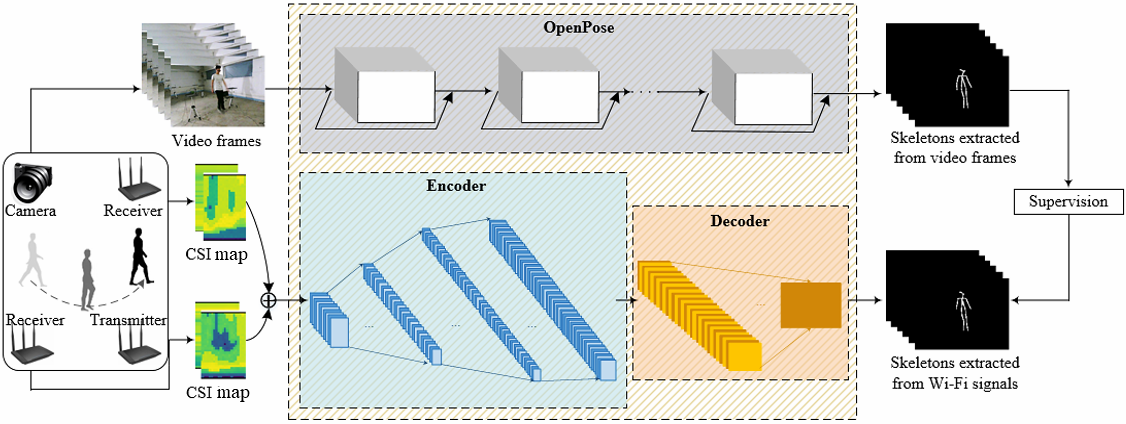}}
\caption{\scriptsize{Network framework for capturing human pose figures. The upper part is mainly for training supervision, while the lower part is for human pose extraction from WiFi signals.}}
\label{fig2}
\end{figure*}

Then, as shown in Figure~\ref{fig2}, we mainly design a neural network to map the CSI maps into human skeleton figures. The encoder network summarizes the information from the original points of view (e.g., the two receivers) and utilizes strided convolutional networks and a SE block \cite{senet2018}. Moreover, the decoder network decodes the poses from the view of the camera by utilizing resize convolutions with nearest neighbor interpolation operation to eliminate the Checkerboard Artifacts \cite{odena2016deconvolution}.

In the process of training, the input of the neural network is \(\left( {{C_1},{C_2}} \right)\), and the output is the predicted human skeleton figure \(P\).
Supervised by \(S\), the human skeleton figure extracted by OpenPose, the neural network is then optimized.

Specifically, the average of binary cross entropy loss for each pixel is applied as the loss function to minimize the difference between the predicted figure and the corresponding annotation:
\begin{equation}
  \mathscr{L} = -\frac{1}{Q}\sum\limits_{i,j} {{S_{i,j}}\log {P_{i,j}} + \left( {1 - {S_{i,j}}} \right)} \log \left( {1 - {P_{i,j}}} \right),
\end{equation}
where \(Q\) refers to the number of pixels in a figure, \({{P_{i,j}}}\) and \({{S_{i,j}}}\) refer to the grayscale values corresponding to the \({\left( {i,j} \right)}\)-th pixel in the figures. It is important to note that we set the total pixels in each figure to \(120 \times 160\) to simplify the network.

\section{Tracking Respiration Status}\label{sec6}
As human respiration does not introduce significant changes on WiFi signals, detecting human respiration is much more challenging, not to mention tracking detailed respiration status. Therefore, we need to design more sophisticated signal processing algorithms to track detailed and accurate human respiration status.

Note that we concentrate on the curves of CSI and utilize the complementarity of CSI amplitude and phase in respiration tracking. So the method we used to extract dynamic component in capturing human pose figures cannot be used here. Thus, in respiration tracking, we remove static component with a Hampel filter. After that, we utilize the periodicity of the respiration status to select the most sensitive signal. To remove the environmental noises, the selected signal is filtered by a wavelet filter. Moreover, we use the peak detection method to detect abnormal respiration. In this section, we will discuss signal processing and abnormal respiration detection respectively.

\begin{figure}[!b]
\centerline{\noindent\makebox[\textwidth] {
    \includegraphics[width=0.42\paperwidth]{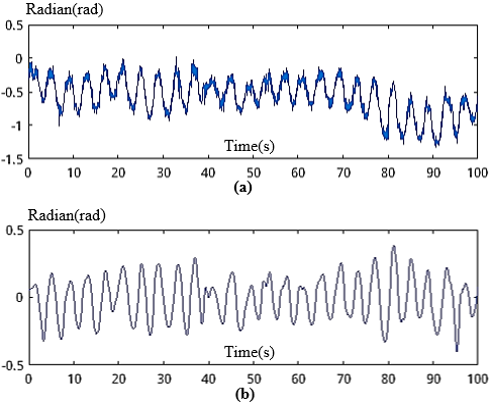}}}
\caption{\scriptsize{Detailed respiratory status curves obtained. (a): A detailed respiratory status curve before wavelet transform. (b): A detailed respiratory status curve after wavelet transform.}}
\label{figfilter}
\end{figure}

\subsection{Signal Processing}
Firstly, a Hampel filter is utilized for extracting and removing the static component described in Equation \ref{S&D path} with a large window size of 100 and a small threshold of 0.01. And the other Hampel filter is used to remove outliers with the window size of 20 and the threshold of 3.

Then we should determine which data stream on which subcarrier contains more respiration information according to CSI amplitude-phase complementarity. We assume that the respiration rate of a human is constant for a short time. Therefore, we define Respiration-to-Noise Ratio (RNR) metric as the ratio of respiratory energy to total energy in the time period, quantifying the periodicity of the respiration of the data stream. The amplitude/phase data among all subcarriers which have the maximum RNR should be considered as the most sensitive data.

Based on the definition above, we find the Fast Fourier Transform (FFT) bin with maximum energy in the range of human respiration, i.e., 10-37 breath/minute \cite{trackingVS}, and compute \(2\times30\) RNR values on amplitude/phase data of 30 subcarriers. We check which data stream has the maximum RNR value, and then we denote it as the selected signal.

To remove the residual noise, we implement ``db4" wavelet decomposition on the selected signal and use only approximation coefficients to reconstruct the respiratory signal \cite{Contactless}. The detailed respiration status curve can be clearly seen as shown in Figure \ref{figfilter}.

\begin{figure}[htpb]
\centerline{\noindent\makebox[\textwidth] {
    \includegraphics[width=0.42\paperwidth]{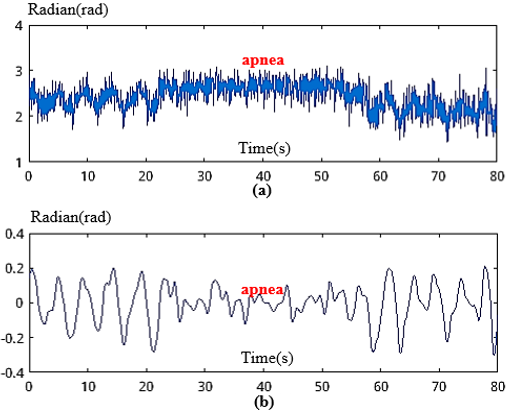}}}
\caption{\scriptsize{The breathing apnea ranges from 25 to 50 seconds. (a): A detailed respiratory status curve  including apnea period before wavelet transform. (b): A detailed respiratory status curve including apnea period after wavelet transform.}}
\label{figAnomaly}
\end{figure}

\subsection{Abnormal Respiration Detection}
Abnormal respiration detection is essential for healthcare monitoring. We suggest the volunteer holding breath for 25 seconds to simulate breathing apnea. In Figure \ref{figAnomaly}, the apnea (no respiration peak) can be roughly observed from the detailed respiration status, However, some subtle fluctuations still remain. Referring to the peak detection method described in \cite{trackingVS}, we applied a minimum amplitude threshold to remove the subtle fluctuations in the period of apnea. If no new peak is detected in the sliding time window, a respiration apnea can be fed back from the system.

\section{Implementation}\label{sec7}
\subsection{WiFi Signals Configuration}
All the transceivers in the experiments are configured to operate at the frequency band of 5\(GHz\) and the bandwidth is set as 20\(MHz\). At the same time, CSI-Tool \cite{halperin2011tool} is installed on all of them. Moreover, the numbers of antennas used by transmitters and receivers are 1 and 3, respectively. To avoid interfering with each other, our human pose capturing transceivers and respiration tracking transceivers work in different WiFi channels, which are set randomly at frequency of 5.28\(GHz\) and 5.68\(GHz\), respectively. It should be noted the other existing WiFi networks, such as campus networks, continue to operate throughout the experiments.

The size of the experimental environment is approximately \(7m \times 8m\). Note that there are two scenarios for capturing human pose figures, including a visible scenario and a through-wall scenario. Visible scenario in our experiments refers to there is no obstacle between any transceiver and the human while through-wall scenario in our experiments refers to all the transceivers are behind walls. Besides, the distance between the respiration tracking transceivers is 1.5\(m\). All the above are consistent with the actual indoor usages, which will be covered in detail in Section \ref{sec9}.

In the process of experimental data collection for capturing human pose figures, we set the collection frequency of CSI as 100\(Hz\), and set the video collection frequency as 20\(Hz\). In other words, for each receiver, every 5 CSI samples correspond to a video frame synchronously via timestamps. Note that given the continuity of human poses over time, 5 synchronized CSI samples and the previous 15 (20 CSI samples in total) are combined to form a CSI map corresponding to a video frame, as described in Section \ref{sec5}.

\subsection{Devices Synchronization}
The Network Time Protocol (NTP) are applied to synchronize the two receivers used for capturing human poses. At the same time, a camera is connected to one of the receivers, and the video frames and CSI samples are synchronized with each other by referring to timestamps. The mean synchronization error of the experiments is less than 1.5\(ms\).

\begin{table}[htpb]\small
\caption{Encoder Network Implementation.}
 \label{network1}
\centering
\begin{tabular}{c|c|c|c|c}
\toprule
Encoder & Input Size & Output Size & Kernel & Stride \\
\hline
layer1 &\(30 \times 20 \times 4\)&\(15 \times 10 \times 8\)&\(3 \times 3\)&\(2 \times 2\)\\
layer2 &\(15 \times 10 \times 8\)&\(15 \times 10 \times 8\)&\(1 \times 1\)&\(1 \times 1\)\\
layer3 &\(15 \times 10 \times 8\)&\(8 \times 5 \times 32\)&\(3 \times 3\)&\(2 \times 2\)\\
layer4 &\(8 \times 5 \times 32\)&\(8 \times 5 \times 32\)&\(1 \times 1\)&\(1 \times 1\)\\
layer5 &\(8 \times 5 \times 32\)&\(4 \times 3 \times 128\)&\(3 \times 3\)&\(2 \times 2\)\\
layer6 &\(4 \times 3 \times 128\)&\(4 \times 3 \times 128\)&\(1 \times 1\)&\(1 \times 1\)\\
SE$^{\rm a}$ &\(4 \times 3 \times 128\)&\(4 \times 3 \times 128\)&---&---\\
FC$^{\rm b}$ &\(4 \times 3 \times 128\)&\(8 \times 10 \times 128\)&---&---\\
\bottomrule
\end{tabular}\\
$^{\rm a}$\textit{\scriptsize{SE refers to the Sequeeze-and-Excitation (SE) block.}}\\
$^{\rm b}$\textit{\scriptsize{FC refers to the fully connected layer.}}
\end{table}

\begin{table}[htpb]\small
\caption{Decoder Network Implementation.}
 \label{network2}
\centering
\begin{tabular}{c|c|c|c|c}
\toprule
Decoder & Input Size & Output Size & Kernel & Stride\\
\hline
layer1 &\(8 \times 10 \times 128\)&\(15 \times 20 \times 64\)&\(1 \times 1\)&\(1 \times 1\)\\
layer2 &\(15 \times 20 \times 64\)&\(15 \times 20 \times 64\)&\(1 \times 1\)&\(1 \times 1\)\\
layer3 &\(15 \times 20 \times 64\)&\(30 \times 40 \times 32\)&\(3 \times 3\)&\(1 \times 1\)\\
layer4 &\(30 \times 40 \times 32\)&\(30 \times 40 \times 32\)&\(3 \times 3\)&\(1 \times 1\)\\
layer5 &\(30 \times 40 \times 32\)&\(60 \times 80 \times 8\)&\(3 \times 3\)&\(1 \times 1\)\\
layer6 &\(60 \times 80 \times 8\)&\(60 \times 80 \times 8\)&\(3 \times 3\)&\(1 \times 1\)\\
layer7 &\(60 \times 80 \times 8\)&\(120 \times 160 \times 1\)&\(3 \times 3\)&\(1 \times 1\)\\
\bottomrule
\end{tabular}
\end{table}

\subsection{Neural Network Implementation}
\paragraph{Encoder Network} A pair of synchronized CSI maps is superimposed and then fed into the encoder network. As shown in Table~\ref{network1}, 6 convolutional layers are utilized in the encoder network to extract features, followed by a fully connected layer to directly convert figures. Additionally, ReLu activation functions are applied to each layer. Note that a SE block \cite{senet2018} is utilized after the last convolution layer in order to extract high-level features.

\paragraph{Decoder Network}
The decoder network utilizes resize convolutions with nearest neighbor interpolation operation and contains 7 layers totally. Table~\ref{network2} illustrates the implementation details.

\paragraph{Training Details}
We implement the network by TensorFlow \cite{abadi2016tensorflow}. The results are trained using Adam optimizer for 10 epochs with a learning rate of 0.001 and a batch size of 16.

\begin{figure}[htpb]
\centerline{\includegraphics[width=0.45\paperwidth]{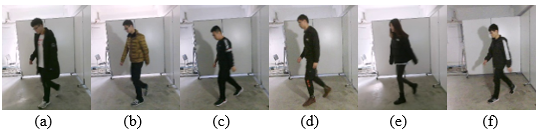}}
\caption{\scriptsize{Volunteers of different clothing, genders, weights and heights in the datasets. (a): The volunteer whose data is untrained and used to further test the generalization of our method in \emph{Case 3} and \emph{Case 4}. (b)-(e): The volunteers whose data is used in \emph{Case 3} and \emph{Case 4}. 75\% of the data is utilized for training the network, and the remaining 25\% for testing the network. (f): The volunteer whose data is used in \emph{Case 1} and \emph{Case 2}. 75\% of the data is utilized for training the network, and the remaining 25\% for testing the network.}}
\label{fig4}
\end{figure}

\subsection{Dataset}
\paragraph{Dataset for Capturing Human Pose figures}
As shown in Figure~\ref{fig4}, the dataset covers 6 volunteers, whose clothing, genders, weights and heights are various. The volunteers are instructed to pose continuously one after another for 12 hours in total. Thus, each receiver collects 4,320,000 CSI samples. At the same time, OpenPose is applied on the synchronized video frames to extract human skeletons to annotate the CSI samples automatically. For each volunteer, half of the data is collected in the visible scenario while the other half is collected in the through-wall scenario.

\paragraph{Dataset for Respiration Tracking}
We conduct extensive experiments with 5 volunteers. The volunteers perform natural breathing in a Fresnel zone region. And the dataset consists of 5 hours of data including 360,000 CSI samples. For comparison purpose, a smartphone accelerometer is attached to the volunteer's chest to record the detailed status of minute movements caused by respiration.

\section{Performance}\label{sec8}
\subsection{Performance of Capturing Human Pose Figures}%穿墙场景
We evaluate the performance of capturing human pose figures in four cases, including:
\begin{itemize}
  \item Case 1:  The same volunteer in the visible scenario.
  \item Case 2:  The same volunteer in the through-wall scenario.
  \item Case 3:  Different volunteers in the visible scenario.
  \item Case 4:  Different volunteers in the through-wall scenario.
\end{itemize}

It is worth emphasizing that there are differences not only among different volunteers, but also among different samples of the same volunteer.

We design the first two cases for users who can accept using a certain amount of time to train models for themselves to get higher accuracy. In these two cases, we use 75\% data of volunteer (f) in Figure~\ref{fig4} to train the network, and the remaining data is used for testing.

Users can also use our model directly without training by themselves. Thus, the generalization ability of the approach needs to be evaluated, that is why we design the last two cases. In these two cases, 75\% data of volunteers (b)-(e) in Figure~\ref{fig4} is used for training the network, and the rest is for testing. At the same time, data from volunteers (a) is excluded from the training data to verify the generalizability of the method we design.

Here we utilize the Percentage of Correct Skeletons (PCS) proposed in our previous work \cite{8943379} to evaluate the performance of our system. It refers to the percentage of Euclidean distance between the binarized annotation and the binarized output not exceeding a certain threshold. It is defined as:
\begin{equation}
PCS \circ \Psi  = \frac{1}{N}\sum\limits_{n = 1}^N {\mathbb{I}\left( {{{\left\| {p_{i,j}^n - g_{i,j}^n} \right\|}_2} \le \Psi } \right)} ,
\end{equation}
where \(\mathbb{I}\) is a logical operation that outputs 1 if true and 0 otherwise. \(N\) is the number of test frames. \({p_{i,j}}\) is the value for the \({\left( {i,j} \right)}\)-th pixel of the predicted figure while \({g_{i,j}}\) is the value for the \({\left( {i,j} \right)}\)-th pixel of the corresponding ground truth figure, where \(i = 1,2, \ldots ,120\) and \(j = 1,2, \ldots ,160\). \(\Psi\) refers to the threshold.

According to \cite{8943379}, with the different values of \( \Psi\), the accuracy and completeness of human poses in the predicted figures, the contrast of the predicted figures and the human positions are different, as shown in Table~\ref{evaluation}.

After that, we evaluate the performance of the four cases respectively as follows.

\begin{table*}[!t]\small
\caption{Levels of Predicted Human Pose Figures Corresponding to Different Values of \( \Psi\).}
\label{evaluation}
\centering
\begin{tabular}{c|cccc|c}
\toprule
\(\Psi\) & Accuracy & Completeness & Contrast & Position & \tabincell{c}{Matching\\ Categories} \\
\midrule
\(25\) &Accurate&Complete&High&Right&---\\
\hline
\(30\) &Accurate&Complete&Lower&Right&\tabincell{c}{Strict Match \\(Matching the \\Whole Pose)}\\
\hline
\(40\) & \tabincell{c}{A Little\\ Inaccurate with\\ Small Limbs}&\tabincell{c}{A Little\\ Incomplete} &\tabincell{c}{A Little\\ Fuzzy}&Right&---\\
\hline
\(50\) & More Inaccurate & More Incomplete & Fuzzier & Right&\tabincell{c}{Loose Match\\ (Matching the \\Core Body)}\\
%\hline
%e&&&&\\
\bottomrule
\end{tabular}
\end{table*}

\begin{table}[htpb]\small
\caption{Results in PCS in Case 1.}
 \label{result1}
\centering
\begin{tabular}{c|c|c|c}
\toprule
\(PCS \circ 25\)&{\(PCS \circ 30\)} & {\(PCS \circ 40\)} & {\(PCS \circ 50\)}\\
\textbf{26.7\%}&\textbf{80.7\%}&98.7\%&\textbf{99.3\%}\\
\hline
\multicolumn{2}{c|}{\emph{Average}$^{\rm a}$} & \multicolumn{2}{c}{27.93}\\
\bottomrule
\end{tabular}\\
$^{\rm a}$\textit{\scriptsize{This is the average of Euclidean distances over all test samples.}}
\end{table}

\begin{table}[htpb]\small
\caption{Results in PCS in Case 2.}
 \label{result2}
\centering
\begin{tabular}{c|c|c|c}
\toprule
{\(PCS \circ 25\)}&{\(PCS \circ 30\)} & {\(PCS \circ 40\)} & {\(PCS \circ 50\)}\\
\textbf{23.4\%}&\textbf{73.0\%}&98.9\%&\textbf{99.4\%}\\
\hline
\multicolumn{2}{c|}{\emph{Average}$^{\rm a}$} & \multicolumn{2}{c}{29.22}\\
\bottomrule
\end{tabular}\\
$^{\rm a}$\textit{\scriptsize{This is the average of Euclidean distances over all test samples.}}

\end{table}

\paragraph{Case 1} Table~\ref{result1} demonstrates the performance of \emph{Case 1}. From the table, we can see that \(PCS \circ 30 = 80.7\%\) and \(PCS \circ 50 = 99.3\%\), which illustrate that 80.7\% of the pose figures captured by our system strictly match the ground truth. And almost all the figures match the ground truth loosely. What's more, the average of Euclidean distances over all test examples is 27.93. We can draw a conclusion that our method can capture both similarities and differences among human poses of a fixed volunteer.

\paragraph{Case 2} Table~\ref{result2} demonstrates the performance of \emph{Case 2}. From the table, we can see that \(PCS \circ 30 = 73.0\%\) and \(PCS \circ 50 = 99.4\%\), which illustrate that 73.0\% of the figures captured by our system strictly match the ground truth. And almost all the figures match the ground truth loosely. And the average of Euclidean distances over all test examples is 29.22.

Compared to \emph{Case 1}, \(PCS \circ 30 \) of \emph{Case 2} drops a little, since the wall may lead to packet loss, while \(PCS \circ 50\) is still considerable. The results show that our system can capture pose figures of a fixed human well even through a wall.

\begin{table}[htpb]\small
\caption{Results in PCS in Case 3.}
 \label{result}
\centering
\begin{tabular}{c|c|cccc|c}
\toprule
{\(PCS \circ \Psi \)}&\emph{(a)} & \emph{(b)} & \emph{(c)}& \emph{(d)} & \emph{(e)}& \emph{Average}$^{\rm a}$\\
\hline
\(\Psi =25\)&\textbf{1.5\%}&\textbf{3.2\%}&\textbf{2.4\%}&\textbf{2.7\%}&\textbf{2.9\%}&\textbf{2.5\%}\\
\(\Psi =30\)&\textbf{20.2\%}&\textbf{26.6\%}&\textbf{24.7\%}&\textbf{28.1\%}&\textbf{34.9\%}&\textbf{26.9\%}\\
% \(\Psi =35\)&56.5\%&55.6\%&43.3\%&55.4\%&58.2\%&53.8\%\\
\(\Psi =40\)&77.3\%&81.9\%&69.3\%&83.7\%&73.6\%&77.2\%\\
\(\Psi =50\)&\textbf{89.2\%}&\textbf{94.6\%}&\textbf{89.3\%}&\textbf{98.4\%}&\textbf{86.6\%}&\textbf{91.6\%}\\
\hline
\emph{Average}$^{\rm b}$&37.57&37.61&39.14&35.13&38.34&37.56\\
\bottomrule
\end{tabular}\\
$^{\rm a}$\textit{\scriptsize{This is the average of \(PCS \circ \Psi \).}}\\
$^{\rm b}$\textit{\scriptsize{This is the average of Euclidean distances over all test samples of a volunteer or all volunteers.}}
\end{table}

\begin{table}[htpb]\small
\caption{Results in PCS in Case 4.}
 \label{result4}
 \centering
 \begin{tabular}{c|c|cccc|c}
 \toprule
 {\(PCS \circ \Psi \)}&\emph{(a)} & \emph{(b)} & \emph{(c)} & \emph{(d)} & \emph{(e)}& \emph{Average}$^{\rm a}$ \\
 \hline
 \(\Psi =25\)&\textbf{0.9\%}&\textbf{2.7\%}&\textbf{1.2\%}&\textbf{1.8\%}&\textbf{1.5\%}&\textbf{1.6\%}\\
 \(\Psi =30\)&\textbf{12.9\%}&\textbf{19.9\%}&\textbf{13.6\%}&\textbf{18.2\%}&\textbf{10.6\%}&\textbf{15.0\%}\\
 % \(\Psi =35\)&56.5\%&55.6\%&43.3\%&55.4\%&58.2\%&53.8\%\\
 \(\Psi =40\)&52.8\%&67.8\%&61.4\%&63.4\%&55.1\%&60.1\%\\
  % \(\Psi =40\)&44.9\%&68.6\%&51.6\%&60.8\%&39.1\%&53.0\%\\
 \(\Psi =50\)&\textbf{72.4\%}&\textbf{89.5\%}&\textbf{83.2\%}&\textbf{86.5\%}&\textbf{70.1\%}&\textbf{80.3\%}\\
 \hline
 \emph{Average}$^{\rm b}$&48.06&39.90&46.68&45.03&48.81&45.70\\
 \bottomrule
 \end{tabular}\\
$^{\rm a}$\textit{\scriptsize{This is the average of \(PCS \circ \Psi \).}}\\
$^{\rm b}$\textit{\scriptsize{This is the average of Euclidean distances over all test samples of a volunteer or all volunteers.}}
 \end{table}
\paragraph{Case 3} Table~\ref{result} demonstrates the performance of \emph{Case 3}. Compared to \emph{Case 1}, the average \(PCS \circ 30 \) drops to 26.9\% while the average \(PCS \circ 50\) drops to 91.6\%. \(PCS \circ 30\) of all the volunteers are much lower because detailed individual features extracted by the neural network are not enough. As \(PCS \circ 50\) of all the volunteers are comparable to that of \emph{Case 1} and the average of Euclidean distances over all test samples is 37.56, we consider the system can capture pose figures of not only trained humans but also untrained humans.

\paragraph{Case 4} Table~\ref{result4} demonstrates the performance of \emph{Case 4}. Compared with \emph{Case 3}, the performance drops, which is consistent with the result of \emph{Case 2}. We notice that \(PCS \circ 50\) of all the volunteers are acceptable and the average of Euclidean distances over all test samples is 45.70, which demonstrates that our system can capture pose figures of different humans well, whether trained or untrained humans in the through-wall scenario.

The experimental results of \emph{Case 3} and \emph{Case 4} show that the system we design and develop is capable of capturing poses of humans with different clothing, genders, weights and heights, which also suggests that the neural network we design is able to capture the similarities and differences among different humans. Furthermore, from the view of WiFi signals, no matter the gender or the type of clothes, WiFi signals will be reflected only by the human bodies. And at the same time, when reflected by the same pose of different humans, WiFi signals will appear some subtle differences and similarities. What needs to be concerned is that the performance on (c) and (e) is relatively low. This is because (c) is underweight and (e) is overweight compared with other volunteers. However, the performance is still within the acceptable range.

Measured by \(PCS \circ 50\), the performance of our system on untrained humans can be compared with that of trained humans, which indicates that our system has the ability to generalize among untrained humans. It is important to note that \(PCS \circ 30\) of untrained humans is relative low is because the neural network lacks details about untrained humans.

\begin{figure*}[htpb]
\centerline{\includegraphics[width=0.64\paperwidth]{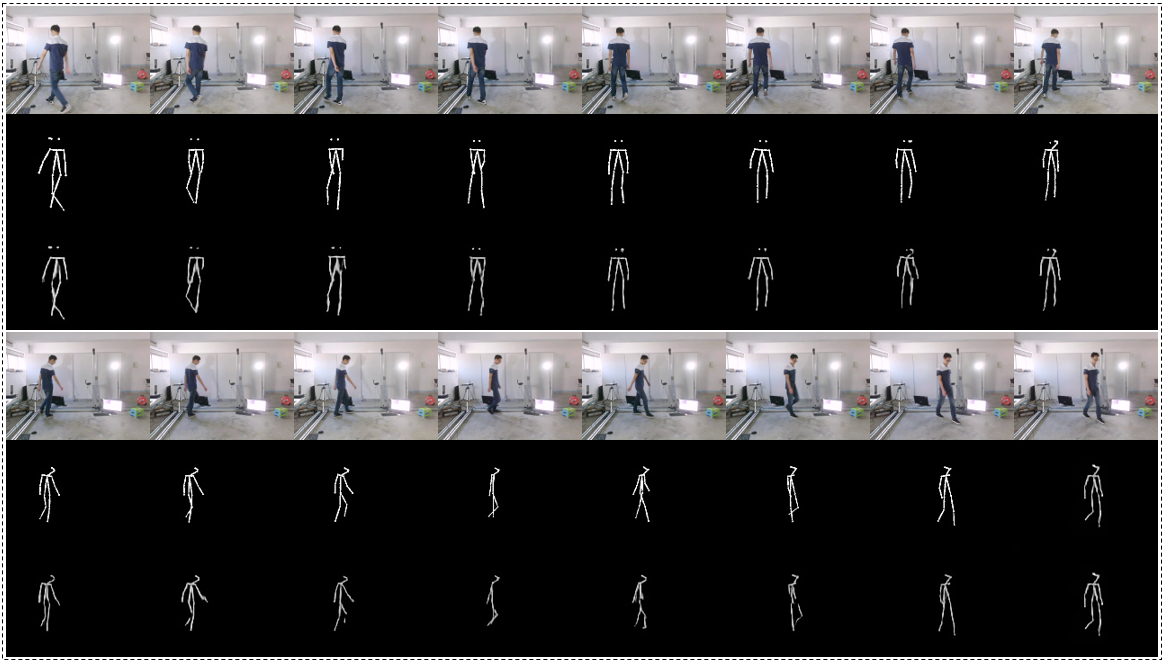}}
\end{figure*}
\begin{figure*}[htpb]
\centerline{\includegraphics[width=0.64\paperwidth]{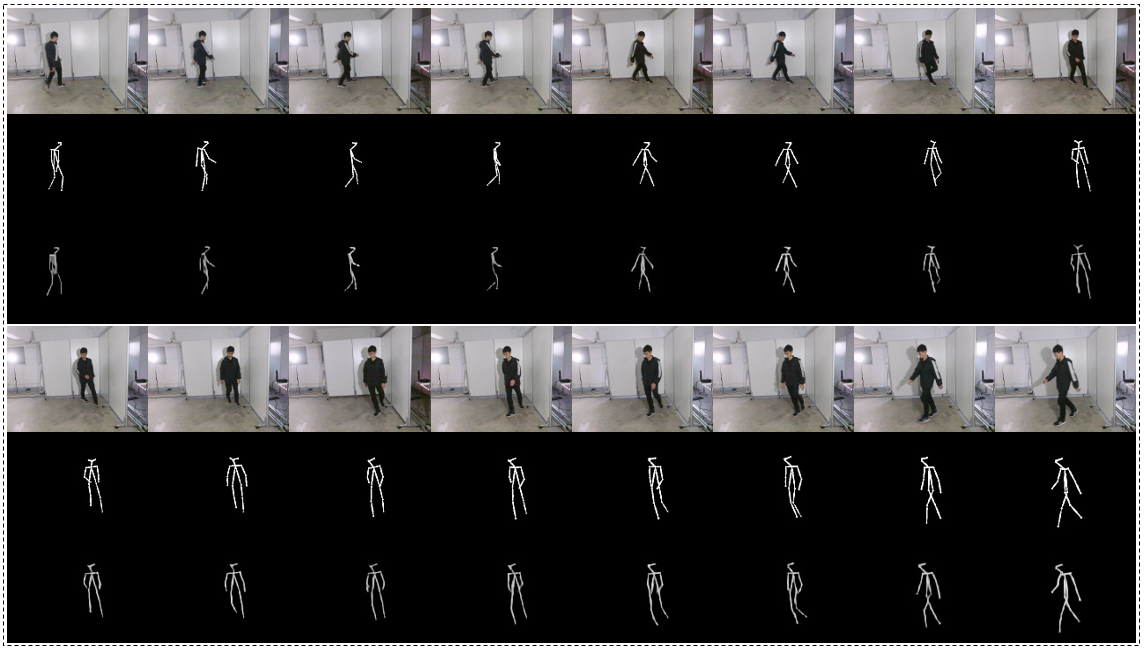}}
\caption{\scriptsize{Test examples which demonstrate that the system we develop continuously tracks the human poses in both the visible scenario and the through-wall scenario. The upper part is in the visible scenario while the bottom part is in the through-wall scenario. In each part, the first and forth pipelines provide human figures which are obtained using the camera, and are shown here for visual reference. The second and fifth pipelines show human skeletons which are extracted by OpenPose, and are shown here as ground truth. The third and last pipelines demonstrate the human skeletons captured by WiFi signals.}}
\label{fig11}
\end{figure*}
The above experimental results illustrate that our system is able to capture human pose figures and achieve the same results as cameras. Yet, unlike cameras, our system can capture human pose figures even through a wall. Further, Figure~\ref{fig11} illustrates test examples of our system capturing pose figures of two volunteers as they walk in different poses and directions continuously in the visible scenario and the through-wall scenario, respectively, which demonstrates that the system we develop is capable of capturing the human pose variability and continuity among different human poses in both visible scenarios and through-wall scenarios.

Consequently, we can then utilize computer vision approaches on the human pose figures captured by WiFi signals to detect abnormal events, i.e., to monitor the elderly in indoor environments.

\subsection{Performance of Tracking Respiration Status}
Volunteers are suggested to hold their breath at will during experiments to imitate the respiration apnea. And the output of the accelerometer is utilized as ground truth to measure the respiration apnea detection performance of our system. Our system achieves 91.2\% detection accuracy.

\begin{figure}[htpb]
\centerline{\includegraphics[width=0.42\paperwidth]{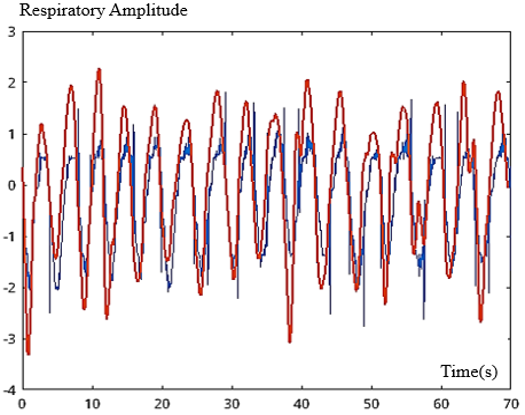}}
\caption{\scriptsize{The comparison between accelerometer data and the detailed respiration status. Red Curve: Accelerometer data. Blue Curve: Detailed respiration status.}}
\label{figcomparison}
\end{figure}
\begin{figure}[htpb]
\centerline{\noindent\makebox[\textwidth] {
    \includegraphics[width=0.42\paperwidth]{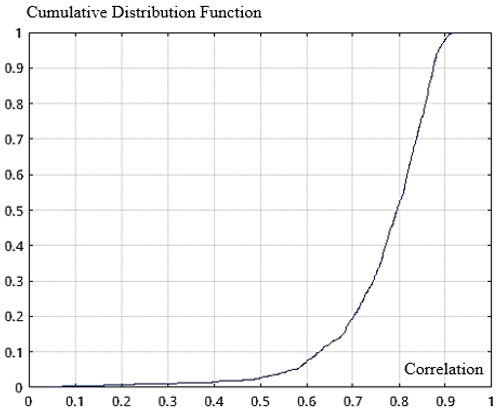}}}
\caption{\scriptsize{CDF of respiration correlation of selected signal and the ground truth.}}
\label{figcorrelation}
\end{figure}

Furthermore, we conduct a similarity comparison by matching the similarity between the ground truth and the respiration status obtained from the selected signal according to the correlation using Pearson correlation coefficient. Figure \ref{figcomparison} shows the comparison between ground truth and the selected signal. Note that the optimal correlation is 1 if the two data streams are entirely consistent theoretically. Figure \ref{figcorrelation} plots the Cumulative Distribution Function (CDF) of the correlation, which shows that the average correlation of our system is 80.0\%. In our view, subtle fluctuations in the abnormal respiratory period would reduce the correlation result.

\section{Discussion}\label{sec9}
Here we discuss some details about the deployment of the approach in practice and possible solutions to the limitations of the current approach.
\subsection{Capturing Human Pose Figures}
\paragraph{Abnormal Events Detection} As mentioned above, our system can capture human pose figures which makes off-the-shelf WiFi devices act as cameras, even through a wall. Then we can use computer vision approaches to detect abnormal events. It should be noted that our system can only capture the poses of a moving human. When the human remains motionless for a long time, the system considers the reflected signal path as a static path and eliminates it through signal processing, so the human will disappear in the figure. This can be considered as the judgment of the human remaining motionless for a long time. When human disappears in the figure during walking, it should be considered that the human leaves the current environment and this does not belong to abnormal events. In addition, in some area, such as on the bed, on the sofa, etc., our system uses the tracking respiration status system as an auxiliary, which makes our system more complete.

\paragraph{Deployment in Different Environments} Under different environmental conditions, the propagation of WiFi signals also varies widely, which is an open problem to deploy WiFi sensing systems in different environments \cite{al2019channel}. Nevertheless, our system removes the static component of CSI, which makes it possible for our system to be deployed in different environments.

\paragraph{Deployment in Through-wall Scenarios}
As mentioned above, WiFi signals have the property to traverse walls. The wall only weakens the signal power received by the receivers, but it does not affect the multipath components associated with human poses, so our system can work in through-wall scenarios accordingly. Therefore, in a typical in-home environment, only one transmitter and two receivers need to be deployed for capturing human pose figures.

\paragraph{Scope and Limitations}
Several limitations exist in our current system:
\begin{itemize}
  \item Small limbs of humans displace, disappear or even overlap in individual locations. The main reason is that the spatial resolution of the off-the-shelf WiFi is insufficient. This could be improved by channel switching and aggregation algorithm proposed in our another work \cite{WiRIM}.
  \item As the camera field-of-view (\(60^\circ \)) is narrower than a WiFi antenna (\(360^\circ \)). It is not enough to annotate from a camera, which may cause some limbs to be obscured. Annotation with more cameras can improve the results.
  \item This work focuses on in-home monitoring for elderly people who live alone, thus the  experimental scenarios only consist one human and we do not consider multiple human scenarios. For multiple human scenarios, we need to expand the dataset containing multiple human poses and retrain the network. And using computer vision methods like Faster R-CNN \cite{ren2015faster}, Mask R-CNN \cite{he2017mask} and Yolo \cite{redmon2016you}, etc. for body segmentation may improve the accuracy of capturing pose figures of multiple humans.
\end{itemize}

\begin{figure}[htpb]
  \centerline{\noindent\makebox[\textwidth] {
      \includegraphics[width=0.415\paperwidth]{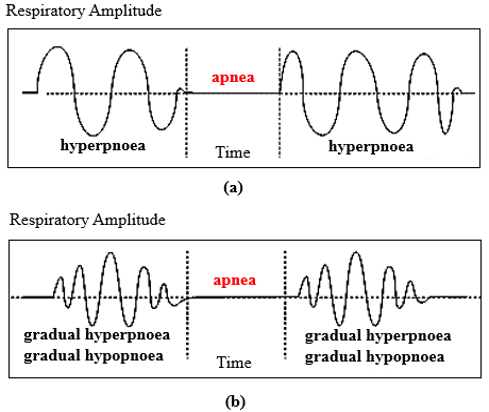}}}
\caption{\scriptsize{Examples of abnormal respiration patterns including apnea periods. (a): Ataxic Respiration. (b): Cheyne--Stokes Respiration.}}
\label{figabnormal}
\end{figure}
\subsection{Tracking Respiration Status}
\paragraph{Abnormal Respiration Patterns Classification}
As shown in Figure \ref{figabnormal}, several abnormal respiration patterns (e.g., Ataxic respiration and Cheyne–-Stokes respiration) include apnea periods. The former is caused by neuron damage and the latter may be caused by damage to respiratory centers or by physiological abnormalities in congestive heart failure. To distinguish different abnormal respiration patterns, we can use some machine learning methods. For example, we can implement Support Vector Machine (SVM) \cite{zhou2016veto} to classify different abnormal respiration patterns: Firstly, we need to extract representative features from the detailed respiration status curves, ranging from common statistics features, such as Max, Min, Mean, Std, etc., to other features like the peak-to-peak interval vectors. Secondly, we can adopt the feature selection method used in our previous work \cite{guo2019wiroi} to obtain sufficient features and reduce complexity. Then using SVM, the system can distinguish different abnormal respiration patterns based on the detailed status curves with apnea periods.

\paragraph{Respiration Tracking Range}
The sensing range is around 1.2 meters away from LoS path when LoS is 1.5m. We think this is sufficient in current scenarios, such as on the bed, on the sofa or in the bathroom, etc.

\paragraph{Scope and Limitations}
We discuss some of the limitations of our system:
\begin{itemize}
    \item To ensure the accuracy of tracking respiration status, we set the distance between the transceivers into 1.5\(m\), which makes the sensing range is around 1.2\(m\). This leads to more WiFi devices being deployed in practice. We will further explore the approach that can enhance the sensing range in our future work for cost savings.
    \item Respiration tracking could be affected by different human poses. To achieve current sensing range, the subject is suggested to face the LoS. By combining capturing human pose figures and pose recognition methods and designing respiration tracking algorithms for different sleep poses, this issue could be addressed.
    \item We focus on in-home monitoring for the elderly who live alone, thus this work only considers one human scenarios instead of multiple human scenarios. For multiple human scenarios, Independent Component Analysis (ICA) \cite{yue2018extracting} and tensor decomposition \cite{wang2017tensorbeat} are worth a try.
\end{itemize}

\section{Conclusion}\label{sec10}
In this work, we present an in-home monitoring approach based on off-the-shelf WiFi devices, which is built on capturing fine-grained human pose figures and tracking detailed respiration status curves. By capturing human pose figures, we can utilize computer vision methods to detect abnormal events. By tracking detailed respiration status curves, we can detect abnormal respiration and even distinguish abnormal respiration patterns for pre-diagnosis. Experimental results show our capturing human pose figure system can achieve comparable performance to cameras and can work well even through a wall, and our tracking respiration status system can efficiently track detailed and accurate respiration status curves. Our work is a completely new way of in-home monitoring which makes all-round daily healthcare information (both behavioral information and physiological information) of the elderly available to caregivers. Compared with other in-home monitoring systems, the proposed approach is low-costs, non-wearable and privacy-protection. Thus, our work can help solving the problem of aging population and the inherent need of assisted-living environments for the elderly to a large extend. Readers can build a more complete system according to Figure \ref{intro} and Figure \ref{fig1}.

%\section{Bibliography styles}

%There are various bibliography styles available. You can select the style of your choice in the preamble of this document. These styles are Elsevier styles based on standard styles like Harvard and Vancouver. Please use Bib\TeX\ to generate your bibliography and include DOIs whenever available.

%\cite{Feynman1963118,Dirac1953888}.

\section*{References}

%\bibliography{mybibfile}
\bibliography{elsarticle-template}

\end{document}